\documentclass{emulateapj}

\slugcomment{Accepted for Publication in ApJ}
\shorttitle{Globular Cluster Enrichment}
\shortauthors{Bailin \&\ Harris}

\newcommand{\Mmin}{{\ensuremath{M_{\mathrm{min}}}}}
\newcommand{\Mmax}{{\ensuremath{M_{\mathrm{max}}}}}
\newcommand{\MSN}{{\ensuremath{M_{\mathrm{SN}}}}}
\newcommand{\MGC}{{\ensuremath{M_{\mathrm{GC}}}}}
\newcommand{\NSN}{{\ensuremath{\bar{N}_{\mathrm{SN}}}}}
\newcommand{\resc}{{\ensuremath{r_{\mathrm{esc}}}}}
\newcommand{\Mretain}{\ensuremath{M^{\mathrm{retain}}}}
\newcommand{\ESN}{{\ensuremath{E_{\mathrm{SN}}}}}
\newcommand{\Msun}{{\ensuremath{M_{\sun}}}}
\newcommand{\Lsun}{{\ensuremath{L_{\sun}}}}
\newcommand{\Zsun}{{\ensuremath{Z_{\sun}}}}
\newcommand{\FeH}{{\ensuremath{\mathrm{[Fe/H]}}}}
\newcommand{\mH}{{\ensuremath{\mathrm{[m/H]}}}}
\newcommand{\Mrad}{\ensuremath{M_{\mathrm{rad}}}}

\begin{document}

\title{Stochastic self-enrichment, pre-enrichment, and the formation
  of globular clusters}

\author{Jeremy Bailin\altaffilmark{1} \and William E. Harris\altaffilmark{2}}
\affil{Department of Physics \& Astronomy, McMaster University,
  1280 Main Street West, Hamilton, ON, L8S 4M1, Canada}
\altaffiltext{1}{bailinj@mcmaster.ca}
\altaffiltext{2}{harris@physics.mcmaster.ca}

\begin{abstract}
We develop a model for stochastic pre-enrichment
and self-enrichment in globular
clusters (GCs) during their formation process.
GCs beginning their formation have an
initial metallicity determined by the pre-enrichment
of their surrounding protocloud, but can also
undergo internal self-enrichment during formation.
Stochastic variations in metallicity arise because
of the finite numbers of supernova. 
We construct an analytic formulation of the combined
effects of pre-enrichment and self-enrichment and 
use Monte Carlo models to verify
that the model accurately encapsulates the
mean metallicity and metallicity spread among
real GCs.  The predicted metallicity spread due to self-enrichment
alone, a robust prediction of the model, is much smaller than the
observed spread among real GCs. This result
rules out self-enrichment as a significant contributor to the metal content
in most GCs, leaving pre-enrichment as the viable
alternative. Self-enrichment can, however, be important
for clusters with masses well above $10^6~\Msun$, which
are massive enough to hold in a significant fraction of
their SN ejecta even without any external pressure confinement.
This transition point corresponds well to the mass at which a mass-metallicity
relationship (``blue tilt'') appears in the metal-poor cluster
sequence in many large galaxies.  We therefore suggest that self-enrichment
is the primary driver for the mass-metallicity relation.
Other predictions from our model are
that the cluster-to-cluster metallicity spread
decreases amongst the highest mass clusters; and that the
red GC sequence should also display a more modest mass-metallicity
trend if it can be traced to similarly high mass.
\end{abstract}

\keywords{globular clusters: general ---
galaxies: abundances --- galaxies: star clusters --- galaxies: formation
--- galaxies: evolution --- methods: analytical}

\section{Introduction}

Globular clusters (GCs) are believed to form in single bursts of star formation
within protocluster clouds, producing bound clusters of
$10^4$ -- $10^7$ stars
which (in most cases) share a single age, metallicity, and element abundance
pattern.  A near-universal feature of the GCs in any one large
galaxy is that they typically follow a {\sl bimodal} metallicity distribution, 
a conclusion now based solidly
on large statistical samples from many galaxies \citep[e.g.][]{larsen01,har06,pen06,pen08}.
The more metal-poor (blue) clusters average $\FeH \sim -1.5$ with a dispersion
$\sigma$[Fe/H] $\simeq 0.25$, and
metal-richer (red) ones average $\FeH \sim -0.3$ with $\sigma$[Fe/H] $\simeq 0.4$
\citep{har06}.

A more recently discovered second-order feature in the metallicity distributions,
that seems particularly to affect the metal-poor component, is a 
{\sl mass-metallicity relation} (MMR):  the blue clusters
become slightly but significantly more metal-rich at progressively higher mass
\citep{har06,str06,mie06}.    No such correlation is seen along
the metal-richer, red GC sequence.
This feature, also colloquially called 
a ``blue tilt'', is as yet poorly understood,
and the various empirical characterizations
of the effect are not yet in mutual agreement, even including the
extreme claim that it is only an artifact of measurement \citep{kun08}.
Some discussions of the MMR in a few individual galaxies claim to find
a systematic correlation of cluster metallicity (color) with
mass (luminosity) that spans the entire GC luminosity range
\citep{str06,spi06,weh08,spi08}.  However, the observational studies based on the biggest
statistical samples (thousands of GCs collected from many
large E galaxies), and carried out with the best photometric measurement
techniques (with careful attention to aperture-size corrections
determined from convolutions of the GC profiles with the point-spread
function of the image) find that the slope of the MMR becomes most noticeable for
the top end of the GC mass range, $M \gtrsim 10^6~\Msun$ 
\citep{har06,mie06,har09}.  For the vast majority of lower-mass clusters,
their mean color stays virtually constant and it is debatable
whether or not any significant MMR trend exists for $M \la 10^6 M_{\odot}$.  

The fact that the MMR becomes prominent only at the high-mass,
high-luminosity end of the GC sequence easily explains why it was
not discovered until recently.  
Massive GCs are rare, and for galaxies with relatively
small GC populations (such as the Milky Way, any dwarfs,
or any non-giant ellipticals),
the GC distribution simply cannot be traced to high enough levels
for any systematic change in cluster color with luminosity to be
noticed.  In the Milky Way,
for example, only $\omega$ Centauri is luminous and massive enough to
be clearly in the ``MMR regime'' \citep{har06}.  To make the picture
additionally puzzling, one or two giant ellipticals seem to have no
discernable MMR at any luminosity level \citep[most notably 
NGC 4472; see][]{str06,mie06} and it is
not out of the question that the amplitude of the effect may differ
from one galaxy to another.

Two other lines of observational evidence have recently emerged to
hint more strongly that the systematic properties of GCs start
to undergo important changes at around the million-Solar-mass point.
One of these is their characteristic radii (usually described by their half-light or
effective radius $r_h$, which is relatively immune to dynamical
evolution).  Empirically, $r_h$ stays relatively uniform at $r_h \simeq 3$ pc for
luminosities $L \la 10^6~\Lsun$ (corresponding roughly to
$M \lesssim 2 \times 10^6 M_{\odot}$), but follows a steeper
relation roughly approximated by $r_h \sim L^{1/2}$ at higher
luminosities \citep{bar07,rej07,ev08}.  
This changeover in their structural-parameter
plane may smoothly link the most massive GCs with other objects
such as UCDs (ultra-compact dwarfs) and the compact nuclei in
dwarf ellipticals.

Lastly, evidence gathered from detailed color-magnitude studies
and abundance measurements of GCs within the Milky Way 
\citep[e.g.][]{bed04,pio07,vil07,mil08,joh08} is now 
showing that the most massive GCs 
(again, those typically above about $10^6~\Msun$)
are often not the clean ``single stellar population'' that
is classically associated with GCs; instead,
they can show two or more distinct sequences in their color-magnitude
diagram that are indicators of more than one generation of stars
with different chemical abundances.

GC formation models to explain and link all these intriguing
characteristics of GCs at different masses are still in 
rudimentary stages.  In this paper, we develop a new quantitative model
to address particularly the metallicity distributions
of GCs and their systematic change with mass.  
Our model assumes that
a globular cluster in the process of formation can be {\sl pre-enriched}
(that is, its protocluster gas will have an initial heavy-element abundance
produced by any generations of stars that happened previously),
and also {\sl self-enriched} (that is, its own first rounds of
supernova can further increase its mean metallicity).  Our model
is intended to explore which of these factors 
will be dominant in a given situation,
and whether the combination is capable of describing the metallicity
distribution for real GCs.

Because all timescales associated with the formation and evolution
of stars decrease strongly with increasing stellar mass, it is possible for the 
first high-mass stars in a protocluster to form, 
live their entire lives, and explode as
supernovae (SNe) while the lower-mass stars are still forming.
The metals produced from high-mass stars within a protocluster may therefore
be incorporated into the lower-mass stars in the same starburst
(self-enrichment).  
In this case, the mean metallicity
of the final cluster will reflect several factors including the initial mass 
function (IMF) of star formation,
the efficiency with which the protocloud gas is turned into stars, and
the ability of protoclusters to hold on to the metals produced by its
SNe (which will be primarily a function of the cloud mass). 
Because the number of SNe in a protocluster may be subject
to small-number statistics, the amount of metal enrichment can be quite
stochastic \citep{cayrel86}, resulting in a metallicity spread that itself
will be a function of cluster mass.

Recently \citet{str08} have published a formation model for GCs also
based on the idea of self-enrichment.  We compare their discussion 
with ours in more detail in a later section below.   We believe that key differences in
the assumptions regarding the protocloud structure and formation
timescales make our model more physically realistic, and as will
be seen below, make it possible to identify the intriguing 
``transition point'' around $10^6~\Msun$ in a more natural way.

\section{Star formation and evolution timescales}%
\label{timescale section}

Previous descriptions of GC and star-cluster formation usually
make the zero-order approximation of an ``instantaneous starburst''
in which the entire stellar IMF springs into place at once.  At some level
this must be an oversimplification:  individual stars have formation
times that depend on their mass, and the starburst epoch as a whole
will be stretched out over a comparable length of time.
Our model of self-enrichment is based instead on the picture
that many or most of the high-mass stars
form, live, die, and eject their newly formed heavy elements
into the surroundings before
the low-mass stars have finished assembling a significant fraction of
their mass.
This scenario, too, must at some level be an idealization, but there is
now significant evidence on both the observational and theoretical
sides to justify it as a useful first-order approximation.

In the context of globular cluster self-enrichment,
 ``high mass'' stars
are those that contribute significantly to the total metal production from
the stellar population while
``low mass'' stars are those that
are still visible today in a Milky Way GC.
The majority of metals are produced in supernovae with progenitor masses of
at least $20~\Msun$ (see Figure~\ref{cumulative metals}), while 
the upper mass limit for ``low mass'' stars relevant here
will be the main-sequence turnoff mass of typical GCs in large
galaxies, $M_{TO} \lesssim 1 M_{\odot}$.
Therefore, the important  \emph{stellar}
timescales are the combined formation and evolution timescale of
stars of at least $20~\Msun$ compared to the formation timescales of
stars of less than $1~\Msun$.

The timescales associated with high-mass star formation are
extremely short. \citet{churchwell02} has shown that the dynamics
of molecular outflows in massive star forming regions
require very high accretion rates that
imply formation timescales of $10^4$ -- $10^5$~yr 
\citep[see also][who infer that the formation times for
massive stars are $< 1$ Myr]{mckee07}.
Once formed, the evolutionary lifetime of high mass stars is
similarly short. \citet{hirschi07} calculates lifetimes of $10$~Myr
for $20~\Msun$ stars down to $< 6$~Myr for stars of
$40~\Msun$ or greater, and \citet{rd05} model combined
effect of metal production and stellar ages and find that most
metals are released within the first $5$~Myr after the first
$100~\Msun$ star dies (see their figure~6).
The timescale for the supernova ejecta to cool is similar,
$\sim 7~\mathrm{Myr}$ \citep{mo77}.
Therefore, even though stars with masses as low as $8~\Msun$ may
explode as supernovae, the key timescale for enrichment is much
shorter than the $\sim 50$~Myr lifetime of those stars.
In short, a reasonable estimate appears to be that the high-mass part of
the IMF can `seed' its surroundings beginning around 5 Myr
after formation.

On the other hand, the formation times for low mass stars are
much longer. The formation timescale should be approximately the
Kelvin-Helmholz time, which rises to $\tau_{KH} \gg 10^7$~yr
for  $M \lesssim 1~\Msun$  \citep[see figure 4 of][]{zy07}.
Therefore, even if all stars in a given starburst begin to form 
simultaneously, the heavy elements from the high mass stars will be
ejected into the protocluster cloud while the low mass stars
are still collapsing. In fact, simulations that couple hydrodynamics
with radiative transfer suggest that feedback from high mass
stars critically shapes the low mass IMF itself \citep{kkm07}, providing
direct evidence that feedback from high mass stars is important
early in the formation of lower mass stars.

These comments still do not include the internal structure of the starburst as
a whole.
The turbulent medium in which star formation occurs is theoretically
expected to result in clumpy star formation, as is seen
observationally in dense star-forming regions today \citep[e.g.][]{fei08,tow06}, 
both in the spatial structure of the protocluster
and in its temporal evolution over the star formation episode.
These studies indicate that the
total durations of starbursts in massive, dense protocluster
clouds such as the ones we are particularly interested in here
are plausibly in the range of $10 - 20$ Myr.  Intervals this long
give the low-mass stars that are formed throughout the burst additional
time to incorporate metals ejected by the higher-mass stars
formed at the onset of the burst. 
While we will not address the effects of this internal
structure on the process of cluster self-enrichment any further, it may be
expected to introduce star-to-star metallicity dispersion within
the cluster; inhomogeneities of this sort have indeed been detected in some
massive GCs (see the references cited above).

\section{Stochastic model of self-enrichment}

We begin by analyzing the simplest case, where the metal content of clusters
comes entirely from self-enrichment, in order to address
whether it can be the dominant contributor. We expand this analysis
to include contributions from both self-enrichment and pre-enrichment
in \S~\ref{fz estimate section}.

\subsection{Mean metallicity}
\label{mean metallicity}
We assume that a globular cluster forms from a molecular cloud of
mass $M_c$, with a star formation efficiency of $f_*$. The stars
are distributed according to an IMF $\xi(m)$ of the form
\begin{equation}
  \xi(m) \equiv \frac{dn}{dm} = A m^\alpha
\end{equation}
between a minimum mass \Mmin\ and a maximum mass \Mmax\ 
(assumed to be $100~\Msun$).
We assume a Salpeter slope of $\alpha=-2.35$, which is an excellent fit
to the observed IMF at $m>1~\Msun$ \citep{chabrier03}.
Although the low mass IMF has
large uncertainties, it is known to turn over and therefore
contain only a small fraction of the total stellar mass.
Our derivation only requires that (a) the high-mass slope of the IMF
and (b) the total fraction of mass contained in high mass stars are correct.
We may therefore assume a Salpeter IMF with an appropriate
low mass cutoff \Mmin\ without loss of generality.
\citet{chabrier03} finds that the mass fraction of stars with
$m > 9~\Msun$ is $0.20$; for our adopted IMF, this implies
$\Mmin=0.30~\Msun$.

The normalization constant $A$ can be found from the condition that
the integrated mass over the IMF is equal to the
total stellar mass \MGC\ ($=f_* M_c$):
\begin{equation}
  A = \frac{f_* M_c (\alpha+2)}{\Mmax^{\alpha+2} - \Mmin^{\alpha+2}}.
\label{a normalization}
\end{equation}

\begin{figure}
\plotone{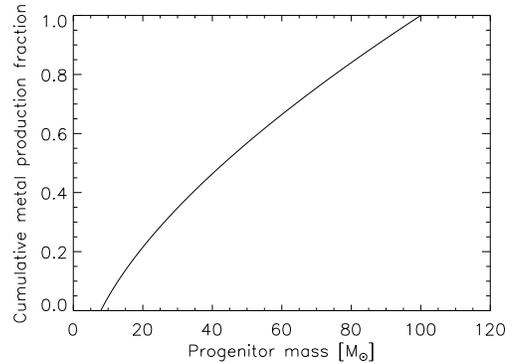}
\caption{\label{cumulative metals}%
Cumulative fraction of metals produced by supernovae with progenitor
star masses less than or equal to the given mass. Almost $80\%$ of
the metals come from stars with initial masses of at least $20~\Msun$.}
\end{figure}

We assume that all stars above a critical threshold $\MSN \sim 8~\Msun$ explode
as SNe.
The mass of metals $m_Z$ ejected by core collapse
supernovae from stars of various progenitor
masses $m$ have been calculated by \citet{ww95} and \citet{nomoto-etal97}.
Assessment of their data shows that they are reasonably 
approximated by the relation
\begin{equation}
  m_Z(m) = (B + C m) m,
\label{WW}
\end{equation}
where $B=1.18\%$, $C=0.548\%$, and all masses are in units of \Msun.
Thus, for example, a $20 M_{\odot}$ star would return to its surroundings
about $2.4 M_{\odot}$
of enriched gas in all heavy elements combined.
We assume that these yields do not depend on the metallicity of
the progenitor stars, an assumption we justify in \S~\ref{yields}.
The total amount of metals produced by SNe within the cloud is therefore
\begin{eqnarray}
  M_Z &=& \int_{\MSN}^{\Mmax} m_Z(m) \xi(m) dm
\\
  &=& A \Biggl[ \frac{B}{\alpha+2}(\Mmax^{\alpha+2}-\MSN^{\alpha+2}) + \nonumber
\\
    & &\frac{C}{\alpha+3}(\Mmax^{\alpha+3}-\MSN^{\alpha+3}) \Biggr].
\label{MZ deriv}
\end{eqnarray}
The relative importance of stars of different mass can
be seen in Figure~\ref{cumulative metals}, where we plot the
cumulative fraction of total metals contributed by supernovae as a function of
the progenitor star mass (i.e.
the plotted value is $f(M) = \int_{\MSN}^{M} m_Z(m) \xi(m) dm /
\int_{\MSN}^{\Mmax} m_Z(m) \xi(m) dm$). Although the IMF rises
to low masses, the yield per star drops almost proportionally and
so stars with masses near \MSN\ do not contribute appreciably
to the total metal production; almost $80\%$ of the metals
come from stars with initial masses of at least $20~\Msun$.

We assume that a fraction $f_Z$ of the metals produced stays within the
the cloud and ends up incorporated in the formation of the lower-mass stars.
For the first stage of our discussion, we assume that $f_Z$ is a constant
free parameter; we will deal with the more realistic case where $f_Z$ depends
on the depth of the potential well in \S~\ref{fz estimate section}.
The total metallicity $Z_c$ of the gas from which the stars are formed is then
\begin{equation}
  Z_c = \frac{f_Z M_Z}{M_c},
\end{equation}
with $M_Z$ taken from equation~(\ref{MZ deriv}).
The resulting metallicity of the low-mass stars that we see in the
cluster today, many Gy later, is then
\begin{equation}
  \log Z_c/\Zsun = 0.38 + \log f_* f_Z,
\label{Zc eqn}
\end{equation}
where we adopt $\Zsun=0.016$ \citep{gre98,van07} although the exact
value is not critical to our argument.
We note that our parameter $f_Z$ is essentially similar to the
``effective yield'' $y_{eff}$ in simple models of chemical
evolution \citep[e.g.][]{pag75,bin98}.

There is no reason
to expect either the star formation efficiency $f_*$ or
the metal retention efficiency $f_Z$ to be near unity. Observations
indicate that $f_* \sim 0.3$ \citep{lad84,lad03,mar08}. Though $f_Z$ is 
less well constrained by direct observation,
we can use the above derivation to constrain the product
$f_* f_Z$. Because the amount of self-enrichment sets only a lower bound to the
total GC metallicity, $\log Z_c/\Zsun$ should 
not be greater than the mean metal-poor GC metallicity of $\mH \simeq -1.25$,
which implies $\log f_* f_Z \le -1.63$ and thus (if $f_* \sim 0.3$)
$f_Z \la 0.08$.

Furthermore, $f_*$ and $f_Z$ may in principle
vary as a function of cloud mass. For example, the fraction of metals that
remain bound to the cloud should increase with the depth of the gravitational
potential well, and therefore $f_Z$ may increase with mass.
This by itself would introduce a positive mass-metallicity relation
in the same sense as the observations listed above. 
We discuss this in more detail in \S~\ref{fz estimate section};
\citet{str08} make an argument that is basically similar.

\subsection{Metallicity spread}
\label{z spread section}

Next, we extend the above derivation to determine the expected 
cluster-to-cluster {\sl dispersion} in the
GC metallicity distribution, due to the stochastic nature of
self-enrichment \citep{cayrel86}.  This will turn out to yield an interesting
constraint on the relative importance of pre- and self-enrichment.

First we note that the mean number of supernovae in a given GC, \NSN, is:
\begin{eqnarray}
  \NSN &=& \int_{\MSN}^{\Mmax} \xi(m) dm
\\
  &=& \frac{A}{\alpha+1} (\Mmax^{\alpha+1}-\MSN^{\alpha+1}).
\label{nsn eq}
\end{eqnarray}

The variance in the mass of metals expelled by a single SN is:
\begin{equation}
  \sigma^2_{m_Z} = \left< m_Z^2 \right> - \left<m_Z\right>^2,
\end{equation}
where
\begin{eqnarray}
  \left< m_Z^2 \right> &=& \frac{ \int_{\MSN}^{\Mmax} m_Z(m)^2 \xi(m) dm }
    {\NSN}
\\
  &=& \frac{A}{\NSN} \Biggl[ \frac{B^2}{\alpha+3}(\Mmax^{\alpha+3}-\MSN^{\alpha+3})
\nonumber\\
   & &+ \frac{2 B C}{\alpha+4}(\Mmax^{\alpha+4}-\MSN^{\alpha+4})
\nonumber\\
   &+& \frac{C^2}{\alpha+5}(\Mmax^{\alpha+5}-\MSN^{\alpha+5}) \Biggr],
\end{eqnarray}
and
\begin{equation}
  \left< m_Z \right>^2 = \frac{M_Z^2}{\NSN^2},
\end{equation}
with $M_Z$ taken from equation~(\ref{MZ deriv}).

Adding together \NSN\ supernovae, the variance in the total metal mass $M_Z$ is
\begin{equation}
  \sigma_{M_Z}^2 = \NSN \sigma^2_{m_Z},
\label{sigmamz2 nsn}
\end{equation}
and the relative scatter in the cloud metallicity $Z_c$ is:
\begin{equation}
  \frac{\sigma_Z}{Z_c} = \frac{\sigma_{M_Z}}{M_Z}.
\label{sigmazz=sigmamzmz}
\end{equation}
The original cloud mass $M_c$ is not observable, but the mass in stars
\MGC\ is.
Combining equations~(\ref{a normalization}), (\ref{MZ deriv}),
and (\ref{nsn eq}) -- (\ref{sigmazz=sigmamzmz}),
we find:
\begin{equation}
  \frac{\sigma_Z}{Z_c} = 0.059 \left(\frac{\MGC}{10^5~\Msun}\right)^{-1/2}.
\label{scatter eqn}
\end{equation}
This formulation reveals a remarkable result: the predicted width of the
metallicity distribution is a function of only the stellar mass in the
globular cluster, independent of the free parameters $f_*$ and $f_Z$.
It is a direct consequence of the fact that higher cluster mass yields
larger numbers of SNe and so reduces the statistical fluctuations
in the enrichment, regardless of how much gas is lost.
This relationship, including normalization, is therefore a
robust testable prediction of the model.

\subsection{Monte Carlo realization}

We next generate a Monte Carlo realization of this process to demonstrate
the stochastic effects more explicitly.
We begin by randomly sampling cloud masses $M_c$ from
a mass function for protocluster molecular clouds of the form
\begin{equation}
  \frac{dN}{dM_c} \propto M_c^{-2}
\label{cloud mass function}
\end{equation}
between a minimum mass $M_0$ and $M \rightarrow \infty$
(\citealt{hp94} estimate a slightly shallower slope for the mass function,
$\sim -1.6$, but our results are entirely insensitive to its value).

For each cloud, we calculate the mean number of stars expected,
$\bar{N}$:
\begin{equation}
  \bar{N} = \int_{\Mmin}^{\Mmax} \xi(m) dm
\end{equation}
\begin{equation}
  = \frac{f_* M_c (\alpha+2)}{(\alpha+1)} \frac{\Mmax^{\alpha+1}-\Mmin^{\alpha+1}}
  {\Mmax^{\alpha+2}-\Mmin^{\alpha+2}},
\end{equation}
and round it to the nearest integer. We then
randomly sample $\bar{N}$ stars from the IMF
and find the stellar mass of the cluster,
\MGC, by taking their sum. Because the
input cloud mass $M_c$ and the mass obtained from a random sampling of the
IMF may not agree, we reassign the cloud mass to $M_c = \MGC / f_*$%
\footnote{We have confirmed that the distribution of reassigned cloud
masses also follows equation~(\ref{cloud mass function}).}. For each star
with mass greater than \MSN, we calculate the mass of metals it produces
using equation~(\ref{WW}), and finally calculate the metallicity of the
cloud by summing up the contributions from each SN and multiplying
by $f_Z / M_c$.

\begin{figure}
\plotone{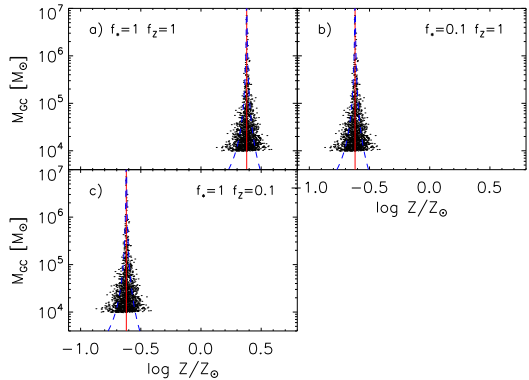}
\caption{\label{zvsm}%
The stellar mass of globular clusters, \MGC\ is plotted against their
metallicity, $\log Z_c/\Zsun$. Panels (a), (b), and (c) correspond
to the models $f_*=1$, $f_Z=1$; $f_*=0.1$, $f_Z=1$; and
$f_*=1$, $f_Z=0.1$ respectively. 
The mean metallicity from equation~(\ref{Zc eqn}) is shown as the
red solid line, and the predicted dispersion from equation~(\ref{scatter eqn})
is shown by the blue dashed lines. Dots indicate the Monte Carlo points.}
\end{figure}

The results of $1500$ Monte Carlo models with a minimum cloud mass
of $M_0=10^4 / f_*~\Msun$
(so that the minimum \MGC\ is constant),
along with the results from the
analytic derivations, are given in Figure~\ref{zvsm}.
It can be seen that the analytic results and the Monte Carlo samples
are completely consistent, and that the free parameters $f_*$ and
$f_Z$ only affect the mean metallicity and not the scatter at a given
globular cluster mass.

\begin{figure}
\plotone{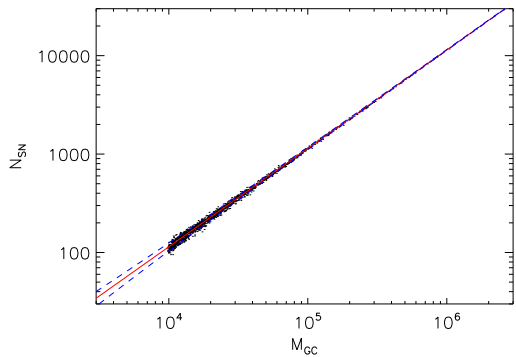}
\caption{\label{nsn plot}%
Number of supernovae per cluster as a function of cluster mass, for
the Monte Carlo realization (dots) and the expected mean from the
equation~(\ref{nsn eq}; red line). The expected Poisson dispersion is
shown as the blue dashed lines.}
\end{figure}

We plot the number of supernovae per cluster in Figure~\ref{nsn plot}.
The average is $\simeq 1$ SN 
for every $100~\Msun$ of stars formed. Thus, for example,
a $10^5~\Msun$ globular cluster -- that is, an object that is thought of as
a classically ``normal'' GC -- will experience on average $1000$ 
supernovae during its formation. 
Note that the spread in cluster
metallicities is not simply the Poisson error of the number
of supernovae, but is also due to the spread in possible metal
contributions of each individual supernova.
For example, the relative Poisson error
on $1000$ supernovae is $\sim 0.03$,
while the actual metallicity spread for clusters of $\MGC=10^5~\Msun$ is
a factor of two larger.

Lastly, we emphasize that the calculations so far assume a 
{\sl constant} gas retention fraction $f_Z$, so that the mean
self-enrichment is independent of cluster mass.  This assumption
is rather obviously unrealistic, since at small enough protocluster
masses the first few supernovae will remove all the gas and fail
to generate any enrichment.  We deal with this feature 
in Section 4 below.

\section{Comparison of self-enrichment model with observations}
\label{obscomp section}
\begin{figure}
\plotone{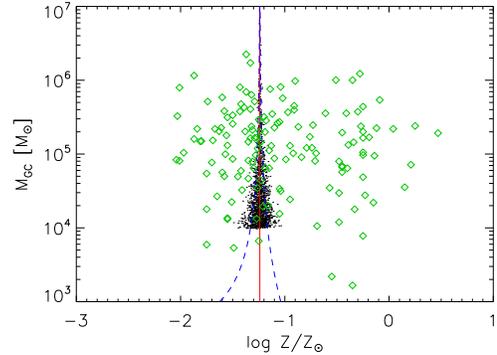}
\caption{\label{observed}%
The red solid line and blue dashed lines show the mean metallicity
and its spread predicted from our pure self-enrichment model for $f_*=0.3$
and $f_Z=0.08$, while the black dots represent 1500 Monte Carlo
realizations. The green diamonds represent the observational data
from \citet{har96} for Galactic globular clusters,
and have been plotted assuming $\log Z/\Zsun = \FeH + 0.25$.}
\end{figure}

In Figure~\ref{observed}, we compare the metallicities and stellar
masses predicted by our self-enrichment model to the observed
Milky Way globular clusters \citep{har96}.
When comparing to observations, we assume
\begin{equation}
  \log Z/\Zsun = \FeH + 0.25,
\end{equation}
with a scatter of $0.1$~dex \citep[e.g.][]{gra89,she01,pri05,kir08}.
We adopt $f_*=0.3$ and $f_Z=0.08$ in order that the predicted
mean metallicity matches the mean metallicity of the observed
metal-poor clusters.

The observed metal-poor clusters have a metallicity spread of
$\sigma \FeH \approx 0.27$
with no obvious dependence on cluster luminosity.
This corresponds to a relative metallicity spread of
$\sigma_Z / Z = \sigma \FeH \> \ln 10 = 0.62$.
In contrast, the self-enrichment model predicts a
relative metallicity spread of only $\sim 0.02$ for clusters
with $\MGC = 10^6~\Msun$, rising to $\sim 0.2$ for very low-mass clusters
with $\MGC = 10^4~\Msun$.
In other words, the observed cluster-to-cluster
metallicity spread is over an order of magnitude larger than
the spread predicted by our self-enrichment model at most
cluster masses.

If self enrichment were the dominant source of metals in the metal-poor
globular cluster population, then the observed and predicted scatter should
be similar, and should be correlated with cluster luminosity.
The dramatic discrepancy between the observed and predicted spread, along
with the absence of any observed relation between cluster luminosity
and metallicity spread,
\textit{strongly suggests that self-enrichment is not the dominant source of 
the heavy elements in most
metal-poor globular clusters}. We conclude that these clusters
must have taken their heavy-element compositions
from the protocluster clouds before star formation began (``pre enrichment''),
and that the observed scatter reflects the variety of environments
in which clusters formed.

\section{The metal retention efficiency and a predicted MMR}
\label{fz estimate section}

In \S~\ref{mean metallicity}, we derived an upper limit of $f_Z \la 0.08$
for the metal retention efficiency based on the mean metallicity of
metal-poor globular clusters. Moreover, the upper boundary
$f_Z \sim 0.08$ corresponds to clusters whose metals are entirely
contributed by self-enrichment, while we concluded 
in \S~\ref{obscomp section} that self-enrichment cannot be the
dominant source of metals.
It is natural to ask whether such low values for $f_Z$ are physically
plausible.
In this section, we develop a simple
energetics argument to obtain a rough estimate for $f_Z$
and explore the consequences of the mass-dependent metal retention
that naturally arises
\citep[see also][who use a similar energetics argument to
examine supernova-driven gas removal from dark matter halo-enshrouded dwarf
galaxies]{ds86,dw03}.

For simplicity,
we assume that the cloud can be approximated by a truncated singular isothermal
sphere (TSIS), with a density distribution of the form:
\begin{equation}
\rho(r) =
\cases{ \frac{M_c}{4\pi r_t^3} \left(\frac{r}{r_t}\right)^{-2} & if $r \le r_t$,\cr
  0 & if $r > r_t$,\cr}
\label{tsis}
\end{equation}
where $r_t$ is the truncation radius.
The potential within the cloud is given by
\begin{equation}
 \Phi(r) = \frac{G M_c}{r_t} \ln r.
\end{equation}

The total kinetic energy imparted into the cluster gas is the sum
of the energy injected by each supernova.
In principle, the important
quantity is the number of supernovae that explode within a cooling time,
and therefore the supernova \textit{rate}, rather than the total number,
is important.
However, in practice the timescale for high-mass star formation is
expected to be $< 1~\mathrm{Myr}$
(see discussion in \S~\ref{timescale section}),
sufficiently shorter than the $\sim 7~\mathrm{Myr}$ in which a supernova
bubble remains hot \citep{mo77} that
all of the cluster supernovae can be approximated as simultaneous.

We next assume that all of the supernova energy is thermalized
within the cluster.
If each supernova imparts an energy \ESN, then the specific kinetic
energy $K$ of the gas after thermalization is
\begin{equation}
  K = \frac{\NSN \ESN}{M_c}.
\end{equation}
If the kinetic plus potential energy of the gas is greater than the
potential energy at the \emph{edge} of the cloud, then the gas can escape.
This occurs for gas beyond an ``escape radius'', \resc:
\begin{equation}
 \Phi(\resc) + K = \Phi(r_t).
\end{equation}
\begin{equation}
 \therefore \resc = r_t e^{-\NSN \ESN r_t / G M_c^2}.
\end{equation}
Because the enclosed mass in a TSIS is proportional to radius,
the fraction of gas within \resc\ is simply $\resc/r_t$. Assuming
that the metals are fully mixed, the metal retention fraction
is equal to the gas retention fraction, and is equal to
\begin{equation}
 f_Z = e^{-\NSN \ESN r_t / G M_c^2}.
\label{fz therm prelim}
\end{equation}
\NSN\ scales with $M_c$, and is well described by
\begin{equation}
  \NSN \approx \frac{\MGC}{10^2~\Msun} = \frac{f_* M_c}{10^2~\Msun}.
\end{equation}
Equation~(\ref{fz therm prelim}) can therefore be recast as
\begin{equation}
  f_Z \approx \exp\left( -\frac{\ESN f_* r_t}{10^2~\Msun G M_c} \right).
\label{fz therm}
\end{equation}

\begin{figure}
\plotone{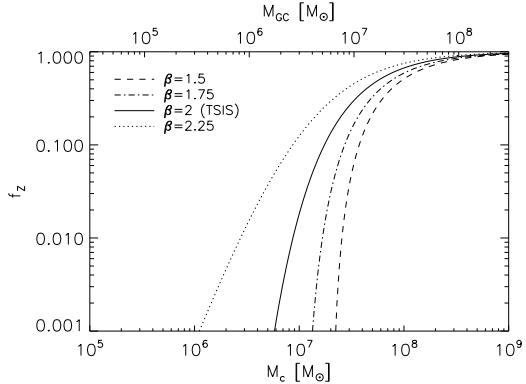}
\caption{\label{fz plot}%
Predicted metal retention efficiency $f_Z$ as a function of protocluster
cloud mass (bottom axis) and final cluster mass (top axis, assuming a
star formation efficiency of $f_*=0.3$).
The solid line is for the fiducial TSIS case, while the other line
styles are the predictions for different protocluster density profiles
(see \S~\ref{density profile section}).}
\end{figure}

In Figure~\ref{fz plot}, we have plotted this relation
assuming $\ESN=10^{51}~\mathrm{erg}$, $f_*=0.3$ and
$r_t=1~\mathrm{pc}$. Note that our adopted radius is smaller than the common
present-day cluster radius of $3~\mathrm{pc}$ due to the expansion
of protoclusters upon removal of their gaseous envelope
\citep{bk07}; if we adopted a larger value for $r_t$, the curve in
Figure~\ref{fz plot} would maintain its shape but
move to proportionally to the right, i.e. to higher mass.
It is apparent that $f_Z \ll 0.08$ for most cluster masses, as required
by our argument that self-enrichment is unimportant.

Another way to express this is in terms of the characteristic cloud mass
at which metals become efficiently retained, $\Mretain_c$, which we define
to be when the argument of the exponential reaches $-1$:
\begin{equation}
 \Mretain_c = \frac{\ESN f_* r_t}{10^2~\Msun G}.
\end{equation}
For the values given, $\Mretain_c = 4 \times 10^7~\Msun$.
In terms of the observable stellar mass,
\begin{equation}
 \Mretain_{GC} = f_* \Mretain_c = \frac{\ESN f_*^2 r_t}{10^2~\Msun G},
\label{mretain gc}
\end{equation}
and has a value of $\Mretain_{GC} = 1 \times 10^7~\Msun$.
This is at the upper range of GC masses, and signals that self-enrichment
may begin to become important for the most massive clusters.\footnote{We note
that the mass \MGC\ as we define it here is the mass just after star formation finishes up.
The observed mass of a GC as we see it \textit{today} is only a lower limit
to \MGC\ because of continual mass loss from
SNe, stellar winds, and ongoing dynamical evaporation of stars
and tidal trimming. For the massive clusters that we are particularly
interested in, this mass loss over several
Gy may be as much as a factor of two \citep{ves03,bk07,wei07}.}

The assumption of complete mixing of the
enriched gas likely results in an overestimate
of $f_Z$ (or equivalently, an underestimate of \Mretain) compared
to the real case where the
cluster wind contains a disproportionate fraction of the metals produced
in the supernovae. The assumption of complete
energy thermalization may either
result in an \emph{overestimate} of $f_Z$ because
it dilutes the supernova energy by the full cluster mass rather than
by a smaller fraction of the material,
or an \emph{underestimate} of $f_Z$ because it assumes that no energy
is lost to radiative cooling.
Using more detailed dynamical considerations, \citet{parmentier-etal99}
argue that protoglobular clouds of significantly lower mass can
contain their supernova ejecta than would be expected from
simple energy considerations (because of their two-phase model
where the enriched gas comes from a previous generation of SNe;
see also \citealp{ml89}),
indicating that our derivation
may underestimate $f_Z$ (overestimate \Mretain).
While the relative magnitudes of
these effects are difficult to assess without full blown hydrodynamic
simulations, their opposite signs reassure us that our results are
unlikely to be incorrect by orders of magnitude.
Given the approximations inherent in our derivation, we caution
against over-interpretation of the detailed form of $f_Z$, but
expect that the general features ---
particularly a transition from no self-enrichment
at low masses to significant self-enrichment above a threshold
GC mass around $\Mretain_{GC}$ --- are robust.

\begin{figure}
\plotone{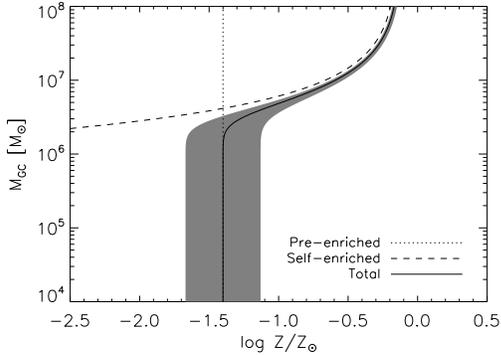}
\caption{\label{mmr plot}%
Predicted globular cluster mass-metallicity relationship assuming a
constant pre-enrichment of $\log Z/\Zsun=-1.4$ and self enrichment
according to our model, with $f_Z$ derived from equation~(\ref{fz therm}).
The dotted, dashed, and solid lines indicate the metallicity due to
pre-enrichment, self-enrichment, and their sum respectively.
The shaded region indicates the predicted scatter, assuming
that the relative scatter in pre-enriched metallicity is
$\sigma \mH = 0.27$.
We have adopted $f_*=0.3$; increasing (decreasing) $f_*$ will
increase (decrease)
the threshold mass at which self-enrichment kicks in by a factor of $f_*^2$
while simultaneously increasing (decreasing) the saturation metallicity
at very large \MGC\ by a factor of $f_*$.}
\end{figure}

In Figure~\ref{mmr plot}, we have plotted the mass-metallicity
relation that would be expected if metal-poor GCs are pre-enriched
to $\langle\FeH\rangle = -1.65$ ($\log Z/\Zsun=-1.4$),
and then further self-enriched according to
our model,
noting again that our adopted supernova yields are independent
of the pre-enrichment level.
The dotted line indicates the pre-enriched metallicity,
the dashed line indicates the metals added by self enrichment,
and the solid line is the total predicted metallicity. For these
sets of parameters, we see that self-enrichment is predicted to
begin contributing at a stellar mass of $\sim 4 \times 10^6~\Msun$,
\emph{very nearly where the observed mass-metallicity relation begins}.
The shaded region indicates the predicted scatter, where we have
assumed that the relative scatter in the pre-enriched metallicities
is $\sigma \mH = 0.27$, as required to fit the observed spread
among Milky Way GCs.
Because the relative scatter due to stochastic self-enrichment
is very small at high masses, the total scatter
is entirely dominated by the scatter in the pre-enrichment values.
Therefore, the metallicity scatter in absolute units is constant,
and the {\sl relative} scatter becomes smaller at higher mass.
Although the details of the mass-metallicity relation are sensitive
to the rough estimate of $f_Z$, and therefore should not be over-interpreted,
it is a clear prediction of our model that the
metallicity \textit{scatter} in dex
should decrease as one goes up the mass-metallicity
relation, preserving an approximately constant absolute scatter $\sigma_Z$.

\begin{figure}
\plotone{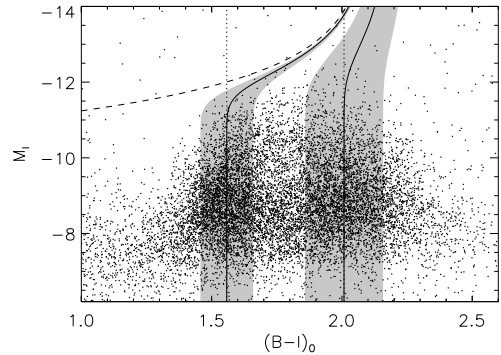}
\caption{\label{elliptical gcs}%
The lines and shaded regions denote the predictions of our
combined pre-enrichment plus self-enrichment model for pre-enrichment
values of $\mH=-1.6$ with scatter $\sigma \mH = 0.27$ (left sequence)
and $\mH=-0.4$ with scatter $\sigma \mH = 0.40$ (right sequence),
translated into the observational color-luminosity space.
Line styles are as in Figure~\ref{mmr plot}. We have overplotted
the observations of giant elliptical GCs from \citet{har09} for
comparison.}
\end{figure}

We directly compare the observed and predicted MMRs in
Figure~\ref{elliptical gcs}, using the data for 6 giant ellipticals
recently remeasured by \citet{har09} from the original data
sample of \citet{har06}. Here, the metal-poor and metal-rich
sequences assume mean pre-enrichments to $\mH=-1.6$ and $\mH=-0.4$
with scatters of $\sigma \mH=0.27$ and $\sigma \mH=0.40$
respectively. These mean metallicities are chosen deliberately
to match the blue and red GC sequences.
We have converted the predicted
metallicities and stellar masses into observed absolute magnitudes
and colors assuming
\begin{equation}
  M_I = 3.94 - 2.5 \log \frac{L}{L_{\odot}},
\end{equation}
\begin{equation}
  (B-I)_0 = 2.158 + 0.375 \log \frac{Z}{\Zsun}
\end{equation}
\citep{har06}.
We adopt
an $I$-band mass-to-light ratio of $M/L_I=1.5~\Msun/L_{\odot}$,
corresponding to $M/L_V=2.0~\Msun/L_{\odot}$ \citep{mclaughlin00}
and a mean GC color $V-I = 1.0$

The model for the metal-poor clusters shows the same general features
as the observations -- a constant metallicity and scatter at most
luminosities, with an MMR for the most luminous clusters. The observed
MMR begins at a lower luminosity than the predicted MMR; this may
be related to our overestimate of \Mretain\ compared to
the more detailed dynamics in \citet{parmentier-etal99},
or to mass loss experienced by these clusters since their
formation, which would result in an overestimate of their mass today
relative to their initial \MGC.

An entirely new feature of our model
is that it predicts the existence of a MMR for the \emph{metal-rich}
sequence of clusters. Because their pre-enrichment level is $\sim 1$ dex
higher, the added factor of self-enrichment has a smaller amplitude
than along the blue sequence and cuts in at a slightly higher
luminosity, but in principle it should be observable.

\begin{figure}
\plotone{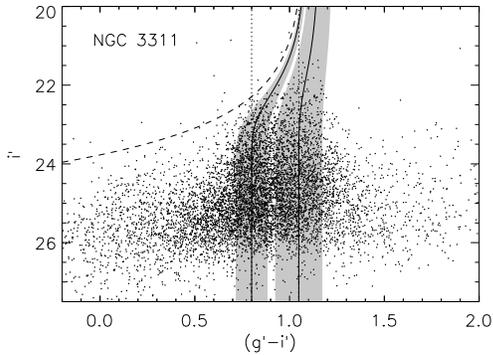}
\caption{\label{ngc3311}%
As in Figure~\ref{elliptical gcs}, but for globular clusters
around NGC~3311 \citep{weh08}.
The model assumes a star formation efficiency
of $f_*=0.2$ and a pre-enrichment level of $\mH=-1.2$ for the
left sequence, with all other parameters identical to the model
in Figure~\ref{elliptical gcs}.}
\end{figure}

To date, we do not have much observational material
for GCSs where the red sequence is seen to extend to high
enough luminosities for its predicted MMR to show up.
However, one such galaxy where hints of this effect can be seen
is the Hydra cD \object{NGC 3311}, shown in Figure~\ref{ngc3311}.
The color-magnitude data \citep{weh08} in $(i', g'-i')$
show the usual presence of the blue and red sequences, but
the red sequence displays an unusually high upward extension
that may connect to the UCD mass regime at $10^7~\Msun$
and above \citep{weh07} and also shows a small `swing' in mean
color further toward the red.
We have overplotted our model predictions,
assuming the following transformations,
\begin{equation}
  i' = 37.90 - 2.5 \log \frac{L}{L_{\odot}},
\end{equation}
\begin{equation}
  g' - i' = 1.173 + 0.311 \log \frac{Z}{\Zsun}
\end{equation}
\citep[based on][]{pen06},
and assuming a star formation efficiency $f_*=0.2$. The slight reduction
in $f_*$ from the fiducial value of $f_*=0.3$ is required in order to fit
the luminosity at which the MMR begins; the sensitive dependence of
$\Mretain_{GC}$ on $f_*$ in equation~(\ref{mretain gc}) means that
very little change is required. In Figure~\ref{ngc3311}, it can be
seen that the strong MMR of the metal-poor population and the
mild MMR of the metal-rich population are both well reproduced
in our model.
We believe that new searches for this MMR effect at the high-mass end
of \emph{both} red and blue sequences would be an extremely effective
test of the basic features of our model. They would best be
looked for in individual supergiant ellipticals with the largest
possible GC samples.

\section{Caveats}

\subsection{Sensitivity to model parameters}

Our numerical results may be influenced by the particular values
we have adopted for
parameters that are either unconstrained or that have significant
uncertainty. We address here the influence that each of these
parameters has on our numerical and qualitative conclusions.

\subsubsection{Initial Mass Function}
The IMF is defined by three parameters: \Mmin, \Mmax, and $\alpha$.
Changes of $20\%$ in \Mmin\ result in changes of $5$--$10\%$
in both the metallicity and its cluster-to-cluster scatter,
while changes of $20\%$ in \Mmax\ result in changes of $\sim 15\%$
to the metallicity and its scatter.

Flattening the slope of the IMF to $\alpha=-2.1$ increases
the mean metallicity by a factor of two, while steepening it
to $\alpha=-2.5$ decreases the mean metallicity by $35\%$; however, the
changes to the scatter are in both cases less than
$25\%$.
Because changes to $\alpha$ result in changes to the mean metallicity
that are comparable to the observed scatter, scatter in $\alpha$ between
clusters could increase the predicted metallicity scatter to the
observed level, assuming that the other parameters of the IMF are
kept constant. However, this is likely a poor assumption: the fraction
of the stellar mass contained in high-mass stars,
which determines the number of supernovae and therefore the
Poisson scatter, is a more robust
quantity than the value of each parameter of the IMF.
We conclude that our qualitative results are
robust to any realistic uncertainty in the IMF.

\subsubsection{Supernova yields}\label{yields}
Because our equation for the heavy-element yields is fit to a small
number of models in \citet{ww95} and \citet{nomoto-etal97}, there
is considerable uncertainty in the parameters. However, our results
are completely insensitive to the $B$ parameter in equation~(\ref{WW}),
and while changes in the $C$ parameter result in proportional changes
to the mean predicted metallicity, they have no effect on the metallicity
scatter.
Another method of effectively changing the supernova yield is to
change the minimum supernova progenitor mass, \MSN. However,
most of the metals are produced by stars with masses
$M_* \gg \MSN$, and so changes in \MSN\ have no noticable effect
on either the mean metallicity or its scatter.

Another potential source of metallicity scatter is from
metallicity-dependent supernova yields. If the mass of metals ejected
from supernovae depended strongly on the metallicity of the
progenitor star, then a small scatter in pre-enrichment
levels could be leveraged into a large scatter in final metallicity.
However, while the yields of individual elements vary significantly
with progenitor metallicity,
the total heavy element yields in the
supernova models of \citet{ww95} are very similar for
models with initial metallicities ranging from $10^{-4}$ to $10^{-2}~\Zsun$,
ruling out this possibility.

\subsubsection{Star formation efficiency}
The predicted metallicity scatter for self enrichment
is entirely independent of the mean value of
the star formation efficiency, $f_*$. Therefore, uncertainty in its
value does not alter our conclusions. However, cloud-to-cloud
\textit{scatter} in the star formation efficiency translates directly
into metallicity scatter \citep[e.g.][]{rd05}.
We therefore cannot rule out a large
mass-independent variation in $f_*$ as an explanation for the
observed metallicity scatter among metal-poor Galactic GCs.
However, our metal retention model of \S~\ref{fz estimate section}
suggests that if self-enrichment were important, there
would be an observable MMR for these clusters
(only clusters with masses $M \gg \Mretain_{GC}$ would exhibit
no MMR, a regime that Galactic GCs clearly do not inhabit),
which is not observed.

\subsubsection{Metal retention efficiency}
\label{caveat:fz}
The metal retention efficiency, $f_Z$,
occupies a similar role as $f_*$:
its mean value has no effect on the predicted metallicity scatter in the
self enrichment model, but cloud-to-cloud \textit{scatter} in $f_Z$
translates directly into metallicity scatter. We therefore cannot
rule out a large mass-independent variation in $f_Z$ as an explanation
for the observed metallicity scatter, but again note that
the MMR that would be expected in the case of self-enrichment is
not observed for the Galactic GCs.

Possible scatter in $f_Z$ due to stochastic effects is discussed
in \S~\ref{stochastic retention}, while scatter in $f_Z$ due to
scatter in cluster radii at a
given mass is discussed in more detail in \S~\ref{rt scatter}.

\subsection{Stellar winds}
\label{stellar wind section}
We have assumed that the global heavy element enrichment and feedback
energetics are dominated by core-collapse supernovae (i.e. Types
Ib and II).
Stellar winds from stars of a wider range of mass contribute significant
amounts of some individual elements, but are not believed to
dominate the global enrichment and are certainly unimportant
energetically in comparison to supernovae \citep[e.g.][]{od08}.
If the metal contribution from stellar winds,
i.e. from more common lower-mass stars,
is higher than currently believed, the metallicity scatter due to
self enrichment would be further reduced from the values we
derive.
Our conclusion that self enrichment produces too little metallicity
spread is therefore robust to the heavy element
contribution from stellar winds.

\subsection{Stochastic metal retention}
\label{stochastic retention}
In the previous picture, all supernovae simultaneously enrich and deposit
energy into the protocluster cloud, which becomes well mixed and
subsequently loses some fraction of
its mass, and so a fraction $f_Z$ of \textit{each} supernova's metals
are retained.
Another possible source of scatter is stochastic metal retention.
In this picture,
the individual supernovae pepper the cloud one at a time, and initially
all metals are retained, but eventually one supernova
injects enough energy that the total remaining gas can be unbound.
Then, without any remaining gaseous envelope, the ejecta from all 
subsequent supernovae can escape freely and the enrichment stops.
In this picture, a fraction $f_Z$ of the supernovae have \textit{all}
of their metals retained, while \textit{no} metals are retained from
the remainder. The effective number of supernovae is then
reduced from \NSN to $f_Z \NSN$, and the
stochastic relative scatter is increased by
a factor of $f_Z^{-1/2}$ (this can
be thought of as a special case of scatter in $f_Z$, which is discussed above).
For $f_Z = 0.08$, the metallicity scatter would be increased by a factor
of $3.5$. This factor is well below the order of magnitude required
to reconcile the observed and predicted metallicity scatter.

\subsection{Protocluster density profile}
\label{density profile section}
The adopted isothermal density profile for the protocluster cloud,
equation~(\ref{tsis}), is as steep a profile as can be realistically
maintained, and is physically well motivated. However, clouds may
exist with slightly shallower density profiles. We therefore briefly
expand the derivation of \S~\ref{fz estimate section} to generic
power-law profiles.

If the density within the truncation radius is given by
\begin{equation}
\rho(r) = \frac{M_c (3-\beta)}{4\pi r_t^3} \left(\frac{r}{r_t}\right)^{-\beta},
\end{equation}
then for $\beta \ne 2$ the metal retention efficiency is
\begin{equation}
 f_Z = \left[ 1 - \frac{\NSN \ESN r_t
    (2-\beta)}{G M_c^2} \right]^{\frac{3-\beta}{2-\beta}}.
\label{fz beta}
\end{equation}
We have plotted this relation for a few slopes in Figure~\ref{fz plot},
in addition to the fiducial $\beta=2$ TSIS case. It is apparent that
although the $\beta=2$ case has a unique functional form, it is not
``critical'' in the sense of showing fundamentally different behaviour;
in all cases,
the metal retention turns over sharply near \Mretain.
The qualitative effect of a shallower density profile
is that a greater fraction of the gas is located near the edge of the cloud
where it can be more easily unbound; $f_Z$ therefore decreases when $\beta$ is
lowered, and shows a sharper transition to the self-enriched regime.

\subsection{Mass-radius relation}
\label{massrad section}
In \S~\ref{fz estimate section} we assumed that all protocluster clouds have
a constant radius independent of mass.
While this is true for observed GCs with $M \lesssim 2 \times 10^6 M_{\odot}$, 
a mass-radius relation
emerges at higher masses, with radii rising from $r_h \simeq 3$ pc at 
$\sim 10^6 M_{\odot}$ to $r_h \simeq 20$ pc at $\sim 3 \times 10^7 M_{\odot}$
\citep[see][]{rej07,bar07,ev08}.  
Assuming that the truncation radius scales similarly as the half-mass radius,
we parametrize the mass-radius relation as follows:
\begin{equation}
  r_t = \cases{ r_{t,0} & if $\MGC < \Mrad$, \cr
    r_{t,0} \left(\frac{\MGC}{\Mrad}\right)^n & if $\MGC \ge \Mrad$,\cr}
\label{massrad eq}
\end{equation}
and examine the consequences on our previous results.
The case $n = 1/2$ corresponds to simple virial equilibrium
and $n=1/3$ to constant density.  The available data, accumulated from
a combination of the most massive known GCs, UCDs, and dE nuclei,
indicate $n \simeq 0.5$ (see the references cited above), which we adopt here.

With this scaling, the metal retention efficiency becomes
\begin{equation}
  f_Z = \cases{ \exp\left(-\frac{\Mretain_{GC,0}}{\MGC}\right)
   & if $\MGC < \Mrad$, \cr
    \exp\left(-\frac{\Mretain_{GC,0}}{\Mrad^n \MGC^{1-n}}\right)
   & if $\MGC \ge \Mrad$,\cr}
\end{equation}
where we define the critical metal retention mass in the absence
of a mass-radius relation $\Mretain_{GC,0}$ as in equation~(\ref{mretain gc}).
The correct critical retention mass
\Mretain\ is unchanged if $\Mrad > \Mretain_{GC,0}$, but otherwise
increases by a factor of
\begin{equation}
  \frac{\Mretain_{GC}}{\Mretain_{GC,0}} = \left(
    \frac{\Mretain_{GC,0}}{\Mrad}\right)^{\frac{n}{1-n}}.
\end{equation}
For $n=1/2$, the exponent in the equation above is just $1$.
For example, if we adopt $\Mrad = 2 \times 10^6~\Msun$ as the
transition to the mass-radius relation, then the critical
retention mass increases by a factor of $\sim 5$ to
$\Mretain_{GC} = 6 \times 10^7~\Msun$.

\begin{figure}
\plottwo{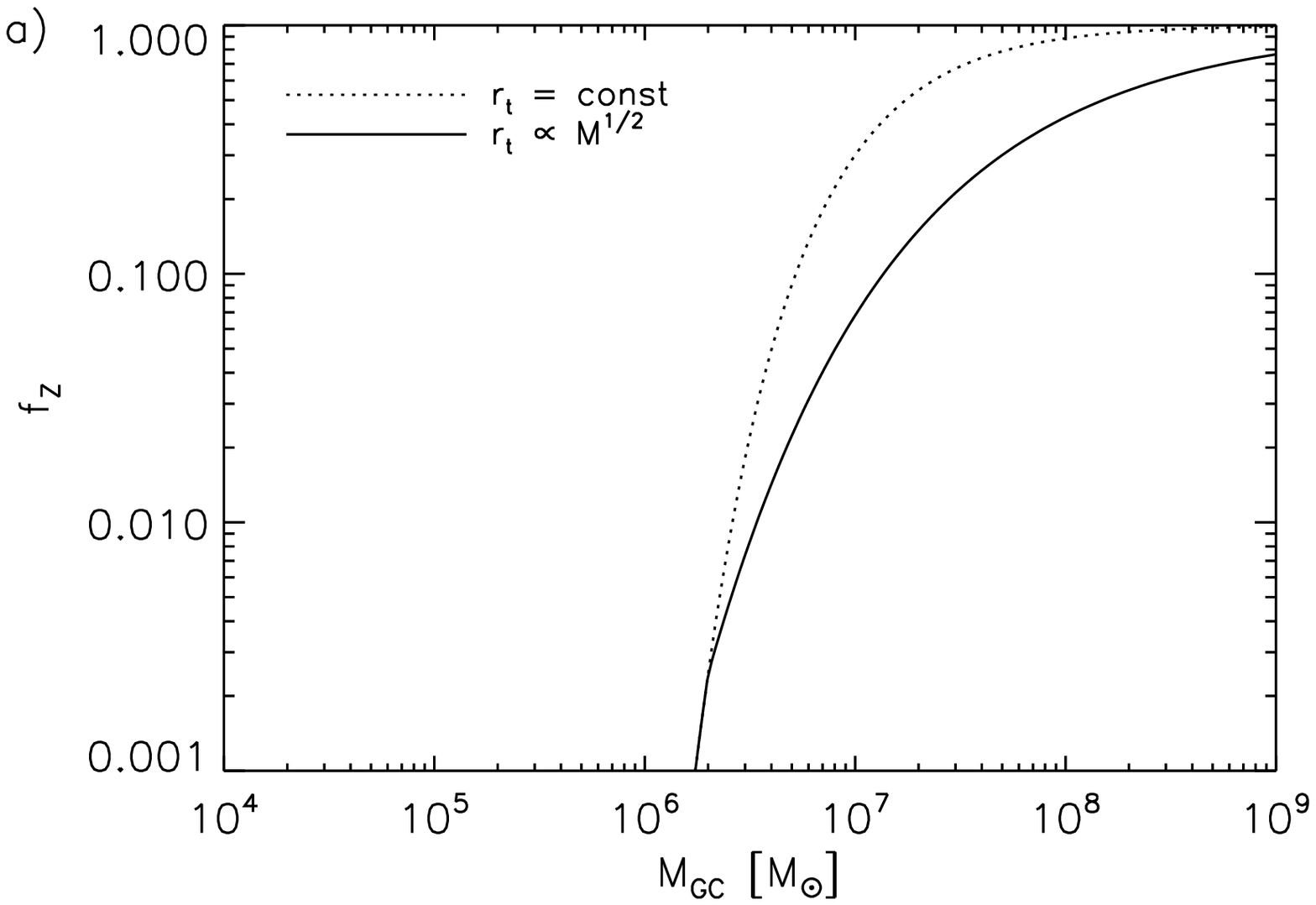}{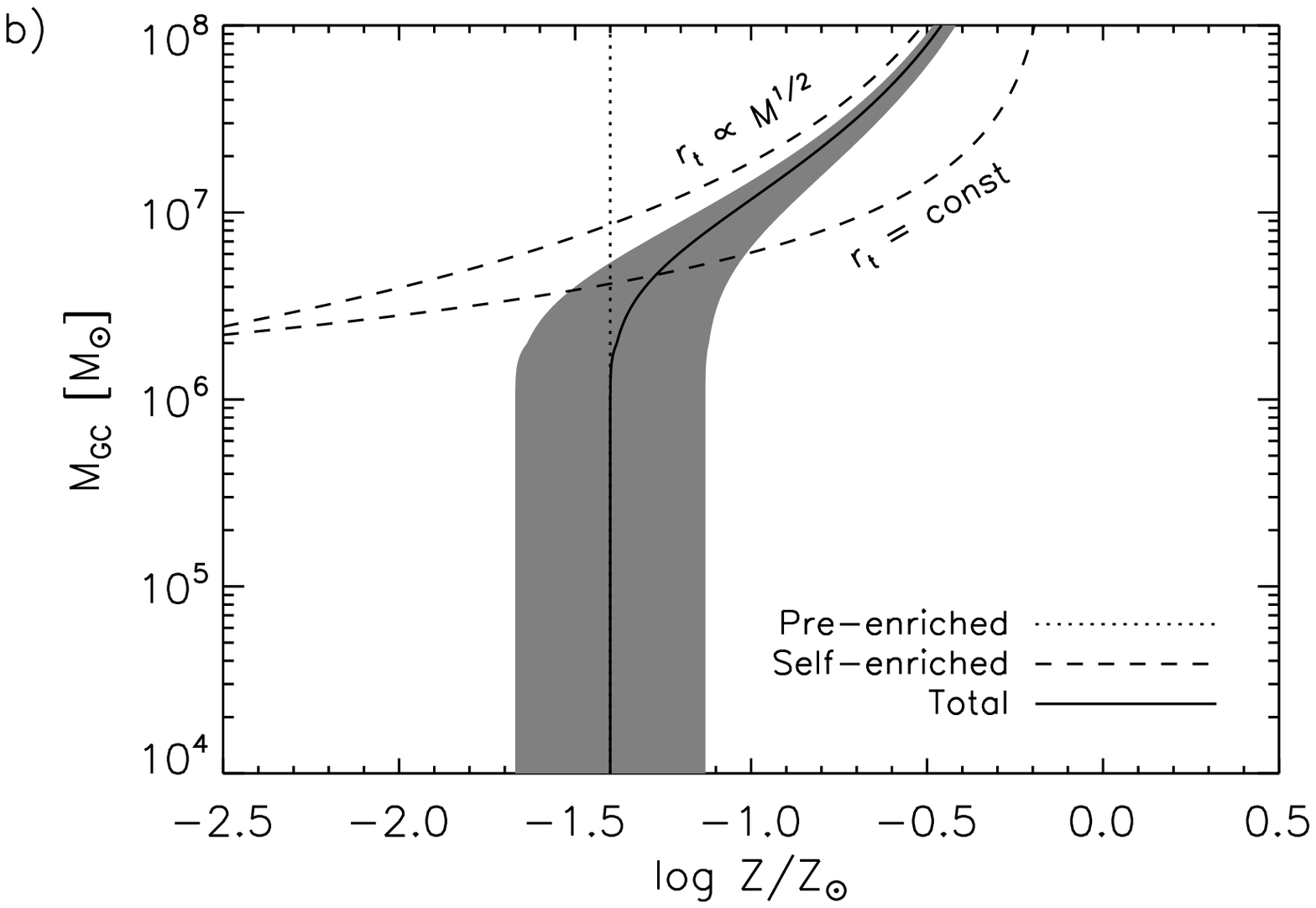}
\caption{\label{massrad}%
(a) Metal retention efficiency, $f_Z$, as a function of cluster mass,
assuming the mass-radius relation of equation~(\ref{massrad eq}) (solid line).
The dotted line indicates the relationship for clusters of constant
radius, as in Figure~\ref{fz plot}.
(b) Predicted mass-metallicity relation, as in Figure~\ref{mmr plot},
but assuming the mass-radius relation of equation~(\ref{massrad eq}).
The self enrichment level for clusters of constant radius
is shown as the lower dashed line.}
\end{figure}

We have plotted our predicted $f_Z$ and mass-metallicity relations
in Figure~\ref{massrad}, assuming the above mass-radius relation
with $n=0.5$ and $\Mrad=2\times 10^6~\Msun$. We have also plotted
the original relations,
where the cluster radius is assumed to be constant,
for comparison. We note that the increased
radius of high-mass clusters reduces their metal retention
because of the shallower potential well, and therefore moves the
scale of self-enrichment to higher mass. However, it also results
in a shallower mass dependence, more similar to the observed
MMRs, and therefore the scale at which self-enrichment
becomes apparent (which is approximately where the dotted and dashed
lines cross in Figure~\ref{massrad}b) is less drastically affected
than the factor of $5$ difference in $\Mretain_{GC}$.

\subsection{Scatter in radius}
\label{rt scatter}
Lastly, we consider the \emph{scatter} in cloud radius at
a given mass and its effect on $f_Z$. The observed scatter in $r_h$
for GCs in the Milky Way is
$\pm0.3$ in $\log r_h$ \citep[with data from][]{har96}; i.e.,
$\sigma_r / r \simeq 0.3$.
As discussed in
\S~\ref{caveat:fz}, $f_Z$ scatter
translates directly into metallicity scatter. We therefore must
determine whether the scatter in $r_t$ can help reconcile the self-enrichment
model with the observed metallicity scatter.

First, we note that the exponential dependence of $f_Z$ on
$r_t$ in equation~(\ref{fz therm}) implies:
\begin{equation}
  \frac{\sigma_{f_Z}}{f_Z} = \frac{\sigma_r}{r_t} \bigl| \ln f_Z \bigr|.
\label{sigma fz vs sigma r}
\end{equation}
From \S~\ref{mean metallicity}, we determined $f_Z \la 0.08$. With
$\sigma_r / r_t \sim 0.3$, the implied scatter on $f_Z$ is
$\sigma_{f_Z} / f_Z \ga 0.75$. This increase in predicted scatter is
sufficient to account for the observed metallicity scatter amongst
low mass GCs.

While this appears to invalidate our argument in \S~\ref{z spread section}
that the magnitude of the observed metallicity scatter rules out self-enrichment
as the dominant mode of metal enrichment, there are critical problems
with this explanation.
First, it is sensitive to the detailed form of $f_Z$, which we do not
claim is robust.
For example, some previous studies of self-enrichment
\citep[e.g.][]{parmentier-etal99}
have modelled the effects of supernovae
as a step-function switch where the protocluster is completely disrupted
below a critical mass, but survives intact with full metal retention
($f_Z=1$) above it;
in such a case, the scatter in $r_t$ would not increase
the metallicity scatter of the extant clusters at all.
Second, the increased scatter is a direct result of the fact that $f_Z$
is a function of the depth of the potential well, which is
proportional to $M_c/r_t$. In other words, it is intimately
tied to the dependence of $f_Z$ on $M_c$ and the
existence of a mass-metallicity relation; however, no
mass-metallicity relation is observed among the low-mass globular clusters.
The only possible ways that this could be masked is if
$M_c \propto r_t$ (in marked contrast
to the observations, which indicate \textit{no} mass-radius relation
in this mass regime), or
if there were a fortuitous cancellation between $f_*$ and $f_Z$
of the form $f_* \propto f_Z^{-1}$, which we consider highly unlikely.

We conclude that the overall picture remains most consistent
with metallicities that are dominated by pre-enrichment for most GCs.

\section{Discussion}

\subsection{Comparisons with Previous Models}

Several previous models for self enrichment have been proposed
in the literature
\citep{cayrel86,ml89,bbt91,parmentier-etal99,rd05,str08}.
Most of these
are based on the picture of \citet{fr85}, in which the protocluster
cloud is a pressure-truncated condensation in the hot halo rather than
having its structure determined by gravity
\citep{bbt91,parmentier-etal99,rd05}. Such models have predicted
no mass-metallicity relation \citep{bbt91}, or a minimal blue tilt that is
much weaker than that observed \citep{rd05}, or an inverse MMR in which
lower-mass clusters are more metal-rich \citep{parmentier-etal99}. These
are therefore not adequate for understanding the metallicities of the
high-mass clusters whose properties are explained by our model.
Most of these models also invoke two distinct star formation episodes:
an initial burst of (possibly massive) primordial composition stars
whose ejecta pollute the protocluster region, followed by a subsequent
burst of star formation from the enriched material. We would classify
such a model,
in which the metals are produced prior to the
burst that creates the stellar cluster we see today,
as ``pre-enrichment''; however, if the delay between
the two bursts is sufficiently short (of order the dynamical time), then
this is a matter of semantics rather than a physical difference
\citep[e.g.][]{cayrel86,rd05}.

In the model of \citet{rd05}, they suggest a number of alternatives
in order to obtain the right cluster-to-cluster
metallicity scatter $\sigma\FeH$, the most notable of which is
to adopt arbitrary and large cluster-to-cluster differences in their
``mixing efficiency'', a parameter resembling our gas retention
fraction $f_Z$ or the effective yield $y_{eff}$. If $f_Z$ is primarily
determined by the initial cluster mass $M_c$, as
we suggest that it should, then we view this
approach as unlikely. Instead, if $\sigma\FeH$ is due to
pre-enrichment, then it can be understood classically as the
result of simple chemical evolution such as a ``leaky box'' model
with a suitable effective yield. For the metal-poor GCs or the
Milky Way halo stars, a reasonable choice is $y_{eff} \simeq 0.03~\Zsun$
\citep{rya01,pra03,van04}.

For our purposes, the most relevant previous
model is that of \citet{str08}, which was
specifically created to explain the observed MMR. Their model,
like ours, invokes
energetic balance between supernova feedback and gravity within
a single episode of star formation to produce an MMR via self enrichment.
They also put forward the ideas that the transition mass marks
the point at which the metals produced by self-enrichment become
comparable to those due to pre-enrichment,
and that the lack of a corresponding ``red tilt'' is due to
the higher level of pre-enrichment among red sequence GCs.
However, there are several key
differences between our models. First, in their model the star formation
efficiency $f_*$ is assumed to vary with mass while the metal retention
$f_Z$ is constant; instead, we argue that $f_Z$ is the physical quantity
much more likely to depend strongly on cluster mass, whereas
$f_* \sim const$.
Second, \citet{str08} derive
power-law scalings for the cluster properties such as mass, metallicity,
and radius, but do not address their amplitude at all; in contrast, our
model is on an absolute scale and
explicitly predicts the transition mass from the pre-enriched regime
to the self-enriched regime.
Third, we predict the existence of a more modest MMR at the high-mass
end of the \emph{red} GC sequence.
Finally, and most importantly,
they examine the mean trends but not
the metallicity scatter between clusters; not only do we explicitly
predict the metallicity scatter due to self enrichment, but we use
this prediction to conclude that self-enrichment is unimportant
among low-mass GCs.

It may be that a combination of ideas from different models
would produce an even more
realistic match to the real GC distributions.
For example, our model predicts a relatively rapid transition
from the pre-enriched regime at low cluster masses to a strongly self-enriched
regime at high masses, which may be more rapid than observed
(note, however, that the mass-radius relation described in
\S~\ref{massrad section} results in a shallower MMR). However, a rapid
transition from no metal retention to significant metal retention,
as in our model, coupled with a star formation efficiency that scales
with mass, as in \citet{str08}, might result in a model that is in
even better agreement with the observations.

Another possible feature to include in a more general model might
be the role of external pressure confinement, which could help
the lower-mass clouds keep back some of their SN ejecta (rather
than having $f_Z \rightarrow 0$ as would be the case if only
self-gravity were operating) and possibly allowing the MMR to
extend to somewhat lower mass.

\subsection{Variation Between Galaxies}
At present, the different sets of observations indicate
that the amplitude (i.e. the exponent in the scaling
$Z_{GC} \propto \MGC^{\alpha}$) of the MMR \emph{may} differ
from galaxy to galaxy. The possibility also exists that
the blue-sequence MMR may set in at different mass
scales and in some cases may be completely absent
\citep{har06,mie06,str06,spi06,spi08,weh08}.
In the context of our simple enrichment-based model, these
differences would boil down to differences in the degree of self-enrichment
at a given cluster mass. The following factors could vary from
galaxy to galaxy and therefore affect the observed cluster
metallicities:

\begin{enumerate}
  \item Mass loss from a cluster depends on the strength of the tidal
	field in which it orbits. Therefore, GCs subjected to
	stronger tidal fields
	should have present-day masses less resembling their original
	\MGC. In this case, the MMR imparted at birth would be weakened or
	destroyed when comparing the present-day masses of the clusters to
	their metallicity. A possible consequence of this effect would
	be that the MMR should be stronger in the outer halos of
	large galaxies.
  \item If the protoclusters were bathed in a particularly strong ultraviolet
	background during formation, (e.g. if the galaxy was undergoing a
	massive starburst during the main epoch of cluster formation),
        the gas may have been less able to cool and therefore the
	star formation efficiency $f_*$ may have been reduced. This would result
	in a lower critical mass \Mretain\ and a lower $Z_c$.
	If $f_*$ were reduced sufficiently,
	the maximum metallicity achievable by self enrichment could even
	drop below the pre-enriched value, resulting in no MMR at all
	\citep[see also the discussion about the influence of a
	UV background in][]{str08}.
\end{enumerate}

\section{Conclusions}
We have developed a model for stochastic self enrichment in
globular clusters. This model predicts both the mean metallicity
and the cluster-to-cluster spread in metallicities that is expected
due to stochastic sampling of the IMF. The predicted metallicity
scatter is an order of magnitude smaller than observed for Galactic
GCs; \textit{this rules out self enrichment as an important contributor
to their global metal content} and leaves pre-enrichment as the dominant
contributor. This conclusion does not depend strongly on the adopted
value of any free parameter and is robust to
reasonable changes to the shape of the IMF, 
the supernova heavy element yields, and the details of how metals
are retained. Although significant cluster-to-cluster scatter
in the star formation efficiency and/or the fraction of metals retained at
a given cluster mass could increase the predicted scatter to the observed
level, they would most likely be accompanied by a significant
mass-metallicity relation at relatively low mass,
which is not observed for the Milky Way.

We have used simple energetics to predict how the metal retention
efficiency, a key parameter in the self-enrichment model,
increases with cluster mass.
We propose a combined pre-enrichment plus
self-enrichment model, where clusters are pre-enriched to the
observed level with significant scatter due to their environment,
and then those clusters sufficiently massive that they can retain a significant
amount of their supernova ejecta are further self-enriched;
the threshold for measurable self-enrichment is predicted to
be typically a few $10^6~\Msun$.
This model matches the main features of the data so far.

The key features of our model can be summarized as follows:
\begin{enumerate}
 \item For globular cluster masses less than $\sim 10^6~\Msun$,
no MMR is expected for either the blue or red clusters; the
mean cluster metallicities should not depend on mass.
 \item Along the blue GC sequence, the MMR should `kick in'
noticeably for \MGC\ higher than a few million \Msun,
reaching $\mH \simeq -1$ at $\MGC \sim 1-2 \times 10^7~\Msun$,
corresponding to the most massive known GCs.
 \item The cluster-to-cluster metallicity spread $\sigma\mH$
remains uniform in the low-mass, pre-enrichment regime, but
decreases at higher mass in the MMR regime where self-enrichment
becomes most important.
 \item A more modest MMR should exist for the red GC sequence,
beginning at a few million \Msun. In at least one cD galaxy
with a very rich GC system (NGC~3311), this effect at the top of the
red sequence may have already been seen.
 \item Within the context of our model, the main free parameter
controlling the level and amplitude of the MMR is the star formation
efficiency $f_*$. Lowering $f_*$ from our baseline value of $0.3$
would reduce the slope of the MMR along both the red and blue GC
sequences and might explain the observed differences between
galaxies.
\end{enumerate}

Two other physical effects not included in our basic model
are a possible correlation of star formation efficiency with cluster
mass, and pressure confinement from the gas outside the protoclusters
at time of formation. Including these effects properly would allow
for a greater range of MMR slopes and onset points and would
be an obvious route to explore in a next stage of studying this
intriguing effect.

\acknowledgments
This work was supported by the Natural Sciences and Engineering
Research Council of Canada through a research grant to WEH.
We are happy to thank Ralph Pudritz and Pamela Klaassen
for useful conversations.

\end{document}